\documentclass[a4paper,reprint,
superscriptaddress,
 amsmath,amssymb,
 aps,
 prc,
floatfix,
longbibliography,
dvipsnames,
accepted=2025-03-20,
twocolumn]{quantumarticle}
\pdfoutput=1
\usepackage[utf8]{inputenc}

\usepackage[main=english,german,ngerman]{babel}

\usepackage{xcolor}
\usepackage{graphicx}
\usepackage{dcolumn}
\usepackage{bm}
\usepackage{siunitx}
\usepackage{dsfont}
\usepackage{amsthm}

\usepackage[ruled,linesnumbered,vlined]{algorithm2e}
\SetKwComment{Comment}{$\triangleright$\ }{}

\let\oldnl\nl
\newcommand{\nlnonumber}{\renewcommand{\nl}{\let\nl\oldnl}}

\usepackage{setspace}

\usepackage{orcidlink}
\usepackage{amsthm}
\newtheorem{problem}{Problem}

\usepackage{ulem}

\usepackage[numbers]{natbib}

\usepackage{mathtools}
\usepackage{ragged2e}

\usepackage{hyperref}

\usepackage{bm}

\usepackage{dsfont}
\usepackage{subcaption}

\usepackage{mathtools}

\usepackage{siunitx}
\usepackage{physics}
\AtBeginDocument{
  \RenewCommandCopy{\qty}{\SI}
}

\usepackage{bbold}
\usepackage{float}
\usepackage{hyphenat}

\usepackage{silence}
\usepackage[shortlabels]{enumitem}

\newtheorem{theorem}{Theorem}

\usepackage[utf8]{inputenc}
\usepackage[toc,page]{appendix}
\usepackage{hyphenat}

\newcommand{\update}[1]{{\color{black} #1}}
\newcommand{\reupdate}[1]{{\color{black} #1}}
\newcommand{\rereupdate}[1]{{\color{black} #1}}

\begin{document}
\title{Early Fault-Tolerant Quantum Algorithms in Practice:\newline Application to Ground-State Energy Estimation}

\author{Oriel~Kiss}
\orcid{0000-0001-7461-3342}
\email{oriel.kiss@unige.ch}
\affiliation{Xanadu, Toronto, ON, M5G2C8, Canada}
\affiliation{European Organization for Nuclear Research (CERN), Geneva 1211, Switzerland}
\affiliation{Department of Nuclear and Particle Physics, University of Geneva, Geneva 1211, Switzerland}
\affiliation{Physics Department, University of Trento, Via Sommarive 14, I-38123 Trento, Italy}

\author{Utkarsh~Azad}
\orcid{0000-0001-7020-0305}
\email{utkarsh@xanadu.ai}
\affiliation{Xanadu, Toronto, ON, M5G2C8, Canada}

\author{Borja~Requena}
\orcid{0000-0003-1904-4137}
\affiliation{Xanadu, Toronto, ON, M5G2C8, Canada}
\affiliation{ICFO -- Institut de Ciències Fotòniques, The Barcelona Institute of Science and Technology, Av. Carl Friedrich Gauss 3, 08860 Castelldefels (Barcelona), Spain}

\author{Alessandro~Roggero}
\orcid{0000-0002-8334-1120}
\affiliation{Physics Department, University of Trento, Via Sommarive 14, I-38123 Trento, Italy}
\affiliation{INFN-TIFPA Trento Institute of Fundamental Physics and Applications, Trento, Italy}

\author{David~Wakeham}
\affiliation{Xanadu, Toronto, ON, M5G2C8, Canada}

\author{Juan~Miguel~Arrazola}
\affiliation{Xanadu, Toronto, ON, M5G2C8, Canada}

\maketitle
\begin{abstract}
We investigate the feasibility of early fault-tolerant quantum algorithms focusing on ground-state energy estimation problems. In particular, we examine the computation of the cumulative distribution function (CDF) of the spectral measure of a Hamiltonian and the identification of its discontinuities. Scaling these methods to larger system sizes reveals three key challenges: the smoothness of the CDF for large supports, the lack of tight lower bounds on the overlap with the true ground state, and the difficulty of preparing high-quality initial states.

To address these challenges, we propose a signal processing approach to find these estimates automatically, in the regime where the quality of the initial state is unknown. Rather than aiming for exact ground-state energy, we advocate for improving classical estimates by targeting the low-energy support of the initial state. Additionally, we provide quantitative resource estimates, demonstrating a constant-factor improvement in the number of samples required to detect a specified change in CDF.

Our numerical experiments, conducted on a 26-qubit fully connected Heisenberg model, leverage a truncated density-matrix renormalization group (DMRG) initial state with a low bond dimension. The results show that the predictions from the quantum algorithm align closely with the DMRG-converged energies at larger bond dimensions while requiring several orders of magnitude fewer samples than theoretical estimates suggest. These findings underscore that CDF-based quantum algorithms are a practical and resource-efficient alternative to quantum phase estimation, particularly in resource-constrained scenarios.
\end{abstract}

\section{Introduction}

Quantum computers may improve the simulation of many-body quantum systems in fields ranging from quantum chemistry to condensed\hyp matter and high\hyp energy physics. While substantial progress has been made towards this goal, there is still a continued need to improve quantum algorithms for a practical impact \cite{kempe2005complexity,lee2023evaluating,liu2022prospects,PNAS_spins, PhysRevA.106.032428} and for solving problems paramount to these systems. One such critical problem is ground-state energy estimation, which historically has been addressed by two primary classes of quantum algorithms: variational methods and quantum phase estimation.
The former class includes variational quantum eigensolvers (VQE) \cite{VQE_preuzzo}, which involve preparing variational quantum states and optimizing their parameters to minimize energy expectation values. Despite its numerous applications \cite{grimsley2019adaptive,  kiss_6li,VQE_QPT,grossi_finite-size_2023,PhysRevA.98.022322,feniou2023overlapadaptvqe,Azad2022,Dumitrescu_2018,PRB_Monaco,perez2023nuclear,Paulin_EF, Sahoo2022}, scaling to larger systems remains challenging due to the substantial sample requirements \cite{Ralli_VQE, Troter_numerics_samples} and trainability issues \cite{ragone2023unified,traps,cerezo2021cost}. In contrast, the quantum phase estimation (QPE) \cite{kitaev1995quantum} algorithm aims to extract the lowest-energy eigenstate by sampling eigenvalues from a distribution induced by an initial state. 

Algorithmically, QPE's strength lies in delivering results with quantitative confidence but experimentally involves challenges such as gate\hyp intensive operations and preparing an initial state with sufficient overlap with the ground state. Consequently, most quantum algorithms for ground-state energies generally align with either noisy intermediate-scale quantum (NISQ) \cite{Preskill2018quantumcomputingin} computing or fault-tolerant quantum computing (FTQC). These observations motivate the exploration of algorithms suitable for early-FTQC devices \cite{EFTQ, early_FTQC} that will still be limited in width and depth but benefit from quantum error correction and mitigation.

Numerous adaptations to QPE have been made to ease implementation on early fault-tolerant quantum computers. Notably, for any system described by a Hamiltonian $\mathcal{H}$ and an initial state $|\Psi\rangle$, we can mitigate the drawback of large circuit depths by analyzing the time series $\langle \Psi|e^{-it\mathcal{H} }|\Psi\rangle$, which only requires one ancilla qubit \cite{somma_2002} and the connectivity induced by the target Hamiltonian. For example, the algorithm proposed by \citet{Lin2022}, inspired by Ref.~\cite{somma2019quantum}, builds the spectral measure's cumulative distribution function (CDF) and identifies the discontinuities with the eigenvalues of the underlying Hamiltonian. In a nutshell, this algorithm, which we shall refer to as the LT algorithm, evaluates the expectation value of the Heaviside function $\Theta(\cdot)$ of the Hamiltonian for an initial state $\langle \Psi|\Theta(\mathcal{H})|\Psi\rangle$, by writing the Heaviside function as a Fourier series and computing the Fourier moments on a quantum computer using the Hadamard test \cite{kiss2024_moments, QA_cleve}.

These methods have been further extended \reupdate{by using an improved Fourier series approximation}~\cite{wan_randomized_2022}, introducing Gaussian kernels \cite{wang_quantum_2022,ding_QMEGS,ding_gaussian}, employing the quantum eigenvalue transform of unitaries (QETU) \cite{QETU}, via rejection sampling \cite{wang2023faster}, by exploring implementation in real quantum hardware \cite{blunt2023statistical, sun2023probing} or against noise models \cite{ding_noise}. \update{Lastly, Ref.~\cite{clinton2024quantum} proposed signal processing techniques for phase-retrieval of time series without using auxiliary qubits, improving the scalability of phase-estimation algorithms. These approaches require prior knowledge of a lower bound on the overlap ($\eta$) between the initial and true ground states to provide performance guarantees. Furthermore, their resource requirements increase quadratically as the quality of the employed initial state deteriorates. 
\reupdate{Navigating such limitations associated with the phase-estimation algorithms would require rephrasing the paradigm of ground-state energy estimation} \cite{v_score_carleo}.
In this work, \reupdate{with this as a motivation,} we adopt a practitioner view, particularly relevant for early fault-tolerant quantum devices: \textit{improving} the energy estimation of \reupdate{an} approximate ground state through classical methods, such as DMRG, rather than striving to solve the exact ground state problem \cite{fomichev2024initial}, since determining the exact ground state is known to be QMA-complete  \cite{gottesman2009quantum, nagaj2008local, cubitt2016complexity, aharonov2009power}. We empirically show in this paper that we can improve the energy estimation, using far fewer resources than anticipated, and \rereupdate{pave the way} towards practical implementation on near-term quantum devices.}

The contributions of this paper are threefold. First, we identify and address the challenges arising when implementing this algorithm in practice. \update{While the implementation of these algorithms is straightforward assuming initial states of good quality, it remains unclear how well they would perform in \reupdate{less optimistic} scenarios. Indeed, preparing good quality initial states is a difficult task in general, \reupdate{particularly for strongly correlated systems and large system sizes.} Hence, in this scenario, the size of the support of the initial state is likely to increase, which }\reupdate{would result in the CDF no longer resembling a series of step functions,} but instead approaches a smooth, monotonically increasing curve. Since identifying discontinuities becomes challenging in this scenario, we propose a different approach, based on signal processing, to find the inflection point of the CDF. \reupdate{We define the inflection point as the lowest energy value with a non-zero probability density, representing the smallest upper bound on the true ground state energy with high confidence.} Concentrating on the inflection point has the added advantage of not requiring a tight estimate of the overlap with the true ground state. \reupdate{While our approach does not differ from the original one \cite{Lin2022} in terms of worst-case guarantees, it provides an automatic way of obtaining ground state energy estimates without requiring a priori knowledge of $\eta$.} Our second contribution consists of quantitative resource estimation for the maximal number of samples required to identify an increase of size $\eta$. \update{More specifically, we compute the prefactors in the estimates from Ref.~\cite{wan_randomized_2022} and empirically compare the required number of samples between the vanilla and our version of the LT algorithm.} Our third major contribution consists of numerical simulations on challenging systems up to $N=26$ qubits. Starting with a density-matrix renormalization group (DMRG) \cite{DMRG_White, DMRG_Wilson} initial state with a low bond dimension and using a low-order Trotter-Suzuki for dynamics, we extract the converged energy of DMRG with larger bond dimension. We also consider a sparsified version of the DMRG initial state as a proxy for an unconverged calculation, which can be potentially more efficient to load onto the quantum computer \cite{sparse_QSP}.  

The paper is organized as follows: we start by describing the LT algorithm in Sec.~\ref{sec:LT}, discuss the practicality of this algorithm in Sec.~\ref{sec:continous}, and introduce our modification involving a procedure to identify inflection points in Sec.~\ref{sec:rpt}. We give the numerical resource estimation in Sec.~\ref{sec:resources} and report the numerical experiments on the capabilities of the LT algorithm in Sec.~\ref{sec:res}. Finally, we conclude with
a discussion of the results and future directions in Sec.~\ref{sec:conc}.

\begin{figure}[!htp]
    \centering
    \includegraphics[width=\linewidth]{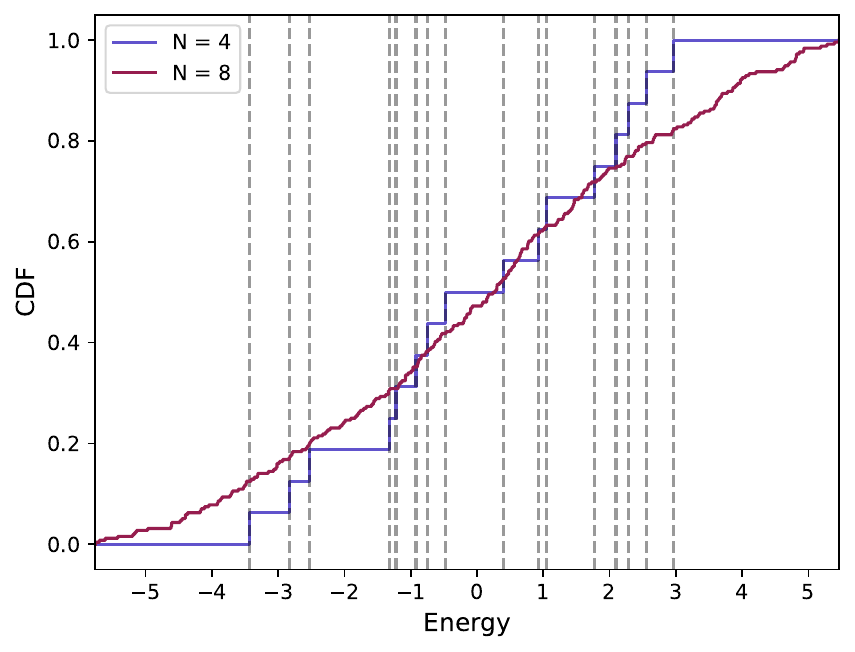}
    \caption{\justifying \textbf{CDF for the XXZ model.} The exact CDFs are shown with coloured-solid lines and show a considerable smoothening behaviour for the eight-spin case. The initial state is chosen at random. The dashed-grey lines show the exact eigenvalues and are being given only for the four-spin case for readability. We note that the eigenvalues are in one-to-one correspondence with the discontinuities, or jumps, in both the CDFs. }
    \label{fig:ex}
\end{figure}

\section{Review of the LT algorithm}
\label{sec:LT}
We shall now describe the LT algorithm in more detail. The purpose of this section is to combine known results in a single place. We start by setting up the problem at hand and then describing the techniques developed in Refs.~\cite{Lin2022, wan_randomized_2022, wang_quantum_2022}. We consider a Hamiltonian $\mathcal{H}$ with the following eigendecomposition
\begin{equation}
    \label{eq:ham_eig_dec}
    \tau \mathcal{H} = \sum_k \lambda_k \Pi_k,
\end{equation}
where $\tau$ is chosen to normalize $\mathcal{H}$ as $\tau||\mathcal{H}||<\pi/2$, and $\Pi_k=|E_k\rangle \langle E_k|$ are the projectors onto eigenspace associated with scaled eigenvalues $\lambda_k = \tau E_k$, ordered as $\lambda_0 \leq \lambda_1 \leq \cdots \leq \lambda_k$. Moreover, we assume access to an initial state $\rho=|\Psi\rangle\langle \Psi|$ and define its spectral measure for $\mathcal{H}$ based on Eq.~\eqref{eq:ham_eig_dec} as
\begin{equation}
    p(x) = \sum_k p_k\tilde{\delta}(x - \lambda_k) = \sum_k \mbox{Tr}[\rho \Pi_k]\tilde{\delta}(x - \lambda_k),
\end{equation}
where $\tilde{\delta}(\cdot)$ is the Dirac delta function and $0 < \eta \leq p_0\equiv |\langle E_0|\Psi\rangle|^2$ defines a lower bound on the ground-state overlap for which we wish to solve the given problem:
\begin{problem}
\label{prob:prob1}
    Given a precision $\delta > 0$ and lower bound on the overlap parameter $\eta > 0$, we seek to decide if
    \begin{equation}
        \label{eq:prob_1}
        \mbox{Tr}[\rho \Pi_{\leq x - \delta}] < \eta\quad\text{or}\quad\mbox{Tr}[\rho \Pi_{\leq x + \delta}] > 0.
    \end{equation}
\end{problem}
In other words, we aim to decide if the ground state energy obeys $\lambda_0 \in [x-\delta,\,x+\delta]$. By answering this question, an approximation of the ground state energy can be found via binary search over an energy grid \cite{Lin2022}. The LT algorithm does this by first approximating the cumulative distribution function (CDF)
\begin{equation}
    C(x) = \underset{i:\lambda_i\leq x}{\sum} p_i,
\end{equation}
of the spectral measure, with error $\epsilon$, with a Fourier series whose moments can be obtained on a quantum computer. In the following, we adopt the choice of Ref.~\cite{wan_randomized_2022} which relates the error to the precision as $\epsilon = \delta/\tau$, which has the advantage of defining a relative precision. We can then extract the eigenvalues of the Hamiltonian, since they appear as discontinuities, or \textit{jumps}, in the CDF, as shown as an example in Fig. \ref{fig:ex} for the periodic four\hyp\ and eight\hyp spins XXZ chain with $J_x=-J_z=1$, using a random initial state. \update{We note that using a random initial \reupdate{state} here emphasizes the challenges for larger systems when high-quality initial states can be challenging to prepare.} The main observation to draw from this example is that it is challenging to identify discontinuities \update{ when the quality of the initial state deteriorates, e.g. for large system sizes;} addressing this regime is a core aspect of this work.

\begin{figure}
    \centering
    \includegraphics[width=\linewidth]{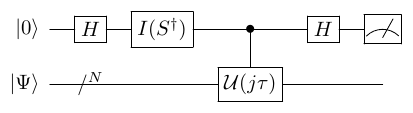}
    \caption{\justifying  \textbf{Hadamard test.} Quantum circuit which can be used to compute the real and imaginary part of the overlap $\langle \Psi|\,\mathcal{U}(j\tau)\,|\Psi\rangle$, where $\mathcal{U}(j\tau)$ is the time evolution operator $\exp{-i\mathcal{H}j\tau}$. The phase gate $S$ is applied on the auxiliary qubit after the first Hadamard $H$ for obtaining the imaginary part. This is adapted from \cite{QA_cleve}.}
    \label{fig:HT}
\end{figure}

The computation of the CDF involves convoluting the Heaviside step function $\Theta(x)$ with the spectral density $p(x) = \sum_k p_k \tilde{\delta}(x-\tau \lambda_k)$, which gives $C(x) = p(x)*\Theta(x)$. This quantity can be computed coherently using quantum signal processing \cite{QSP, QETU}. However, for the early fault-tolerant regime, we instead compute its approximation as a Fourier series by evaluating Fourier moments on the quantum computer and then adding them, weighted by the Fourier coefficients, on a classical device. 
To this end, we first approximate the smoothed Heaviside function as a Fourier series,
\begin{equation}
\label{eq:fourier-series}
F(x;\beta) = \sum_{|k|\leq D} F_{k}(\beta) e^{ikx},
\end{equation}
\reupdate{where the number of Fourier moments (i.e., the runtime of the algorithm) is $D$, the convergence parameter is $\beta$, and the Fourier coefficients $F_k(\beta)$ are} \cite{wan_randomized_2022}
\begin{equation}
\label{eq:fourier_coeffs1}
\begin{split}
    F_0(\beta) &= 1/2, \\
    F_{2j+1}(\beta) &=  -i \sqrt{\frac{\beta}{2\pi}} e^{-\beta} \frac{I_j(\beta) + I_{j+1}(\beta)}{2j+1},\ \text{and}\\ 
    F_{2d+1}(\beta)&= -i \sqrt{\frac{\beta}{2\pi}} e^{-\beta} \frac{I_d(\beta)}{2d+1},
\end{split} 
\end{equation}

with $d$ being related to the runtime as $2d+1=D$ and $I_n(\beta)$ being the $n$-th modified Bessel function of the first kind. This is based on the Chebsyshev approximation of the scaled and shifted error function, which converges to the Heaviside function based on the values of $\beta$ and $D$ ~\cite{wan_randomized_2022}. In order to guarantee an approximation error $\epsilon$, one can take
\begin{equation}
\label{eq:beta}
    \beta = \max{ \left[1,\, \frac{1}{4\sin^2{\delta}} W_0\left(\frac{2}{\pi \epsilon^2}\right)\right]} = \update{\mathcal{O}(\delta^{-2}\log{\epsilon^{-1}})},
\end{equation}
where $W_0(\cdot)$ is the principal branch of the Lambert-W function, together with a number of terms scaling as
\begin{equation}
D=\mathcal{O}\left(\delta^{-1}\log\left({\epsilon^{-1}}\right)\right)\,
\end{equation}
\update{up to a logarithmic factor.}
We guide the reader to \cite[Appendix 1]{wan_randomized_2022} for additional details and an explicit construction.
Using these coefficients, an approximate periodic CDF can be expressed as 
\begin{equation}
\begin{split}
    \tilde{C}(x) &=  \int_{-\pi/2}^{\pi/2}p(y)F(x-y)dy \\
    &= \sum_{|k|\leq D} F_k e^{ikx} \langle \Psi|e^{-i\tau k \mathcal{H}}|\Psi \rangle,
\end{split}
\end{equation}
\update{and we can define the $g_k$ as the Fourier moments 
\begin{equation}
\label{eq:moments}
g_j = \sum_k p_k e^{-iE_j\tau j} = \langle \Psi|e^{-iH\tau j}|\Psi\rangle,
\end{equation} 
whose real and imaginary parts can be computed with a Hadamard test, visualized in Fig.~\ref{fig:HT}.}

Noting that $g_{-k} = g_k^*$,  $F_{2k}=0$, $F_{k} = -i|F_k|$ for $k>0$ and $F_{k} = i|F_k|$ for $k<0$, the approximate CDF (ACDF) can be written as \reupdate{following with $j\equiv 2k+1$}
\begin{equation}
\label{eq:acdf_def}
\begin{split}
\tilde{C}(x)  &=\frac{1}{2} + 2\sum_{k=1}^d |F_{j}|\big( \Re[g_{j}] \sin{jx} + \Im[g_j]\cos{jx} \big).
 \end{split}
\end{equation}
We use $\mathcal{U}(j\tau)=\exp{-i\mathcal{H}j\tau}$ to denote the time evolution operator and define the Hadamard gate $H$ and phase gate $S$ as
\begin{equation}
  H = \frac{1}{\sqrt{2}} \begin{pmatrix}1&1\\1&-1 \end{pmatrix}, \quad  S =\begin{pmatrix} 1&0\\0&i \end{pmatrix}.
\end{equation} 
We note that the computation of Fourier moments can be performed without a direct control on the evolution operator $\mathcal{U}$, e.g., by using control reversal gates \cite{Lin2022, QETU}, which turns out to be always economical compared to the standard approach \cite{kiss2024_moments}. Moreover, such gates also enjoy fast forwarding by a factor of two, thus dividing the maximal circuit's depth by the same amount.

\begin{figure*}[!htp]
    \centering
    \includegraphics[width=\textwidth]{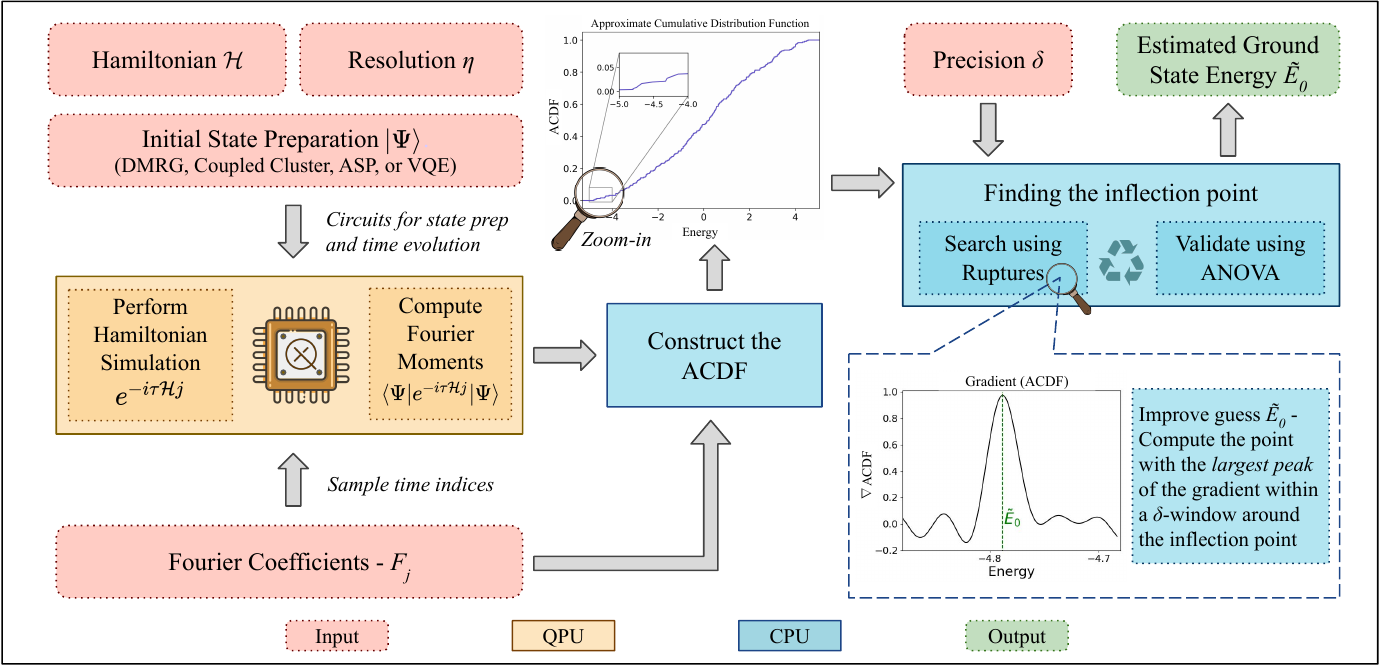}
    \caption{\justifying \textbf{Workflow of the algorithm}: The procedure takes as inputs the Hamiltonian $\mathcal{H}$, initial state $|\psi\rangle$ and tolerance parameters $\delta$ and $\eta$. The Fourier moments, see Eq.~\eqref{eq:moments}, are computed on the quantum computer with time indices $j$ sampled based on the Fourier coefficients of the approximate error function. The approximated CDF is built by summing the moments with the coefficients, as shown in Eq.~\eqref{eq:acdf_def}. Finally, the ground state energy $\tilde{E}_0$ is predicted by first finding the inflection point using \textit{ruptures}, validated by the ANOVA scheme, as an initial guess and improving upon it with the point with the maximal gradient in its $\delta$-neighbourhood (Sec. \ref{sec:rpt}).}
    \label{fig:algo}
\end{figure*}

Finally, to bring the convergence from $\mathcal{O}(\epsilon^{-2})$ down to $\mathcal{O}(\epsilon^{-1})$, and therefore reach the Heisenberg scaling limit, the sum can be computed using importance sampling \cite{Lin2022}. More precisely, \update{we sample a set of $M$ integers $\{k_1, \dots, k_{M} \}$, with each $k_i \in \{1,2, \dots, d\}$.}\reupdate{The probability of observing a $k_i$ is equal to the normalized Fourier coefficients $|F_{k_i}|/\mathcal{F}$, where $\mathcal{F} = \sum_{k=1}^d |F_{2k+1}|$ is the normalization factor.} \update{The corresponding Fourier moments $g_k$ are then measured once and averaged together to build the ACDF as}

\begin{equation}
\begin{split}
\label{eq:gofx}
G(x) = \frac{1}{2} + \frac{2\mathcal{F}}{M}\sum_{i=1}^{M}\Big[ &\Re[g_{k_i}(\tau)] \sin{(k_ix)} \\
+ &\Im[g_{k_i}(\tau)]\cos{(k_ix)} \Big].
\end{split}
\end{equation}

\reupdate{In the context of our approach, this $M$ will correspond to the algorithm's total sample complexity (i.e., the number of shots).} Following the same steps as in~\cite{Lin2022}, one can easily show that the variance of $G$ is bounded by $2\mathcal{F}^2/M$. Once an estimate $\tilde{E}$ is found, we can refine it by computing the derivative of the ACDF and take the maximum on an interval $[\tilde{E}-\delta,\tilde{E}+\delta]$ of the CDF's derivative \cite{blunt2023statistical}

\begin{equation}
\begin{split}
    G'(x) = \frac{2\mathcal{F}}{M}\sum_{i=1}^{M}k_i \Big[ &\Re[g_{k_i}(\tau)] \cos{k_ix} \\
     -& \Im[g_{k_i}(\tau) ]\sin{k_ix} \Big].
    \end{split}
\end{equation}

We note that this is similar in spirit to identifying the maxima of a Gaussian kernel placed around the energy guess $\tilde{E}$~\cite{wang_quantum_2022}. An overview of the whole algorithm is shown in Fig.~\ref{fig:algo}, with input parameters, reconstruction of ACDF and the detection of inflection point, which we cover in greater detail in the next section.  
\section{The LT algorithm in practice}
\label{sec:continous}
The LT algorithm is a candidate ground-state energy estimation algorithm for early fault-tolerant quantum computing (FTQC) devices. Even though it is proven that this algorithm can find the ground-state energy using $M=\mathcal{O}(1/\eta^2)$ samples and maximal depth $D=\mathcal{O}(\delta^{-1}\log{\delta^{-1}\eta^{-1}})$ \cite{Lin2022}, several challenges must be overcome for it to be useful in realistic settings. In this section, we identify these practical problems and propose implementable solutions. Most of them arise from the assumption that we have access to an initial state with at least $\eta$ overlap with the actual ground state, with $\eta$ large enough. However, this is not realistic in general for the following reasons: (i) it is difficult to prepare initial states with significant overlaps, and (ii) there is a lack of techniques to estimate tight bound $\eta$ on the overlap parameter $p_0$.

\subsection{Initial state preparation}
The quality of the initial state significantly constrains the efficacy of quantum phase estimation (QPE) and its variants, specifically its overlap with the true ground state. Therefore, the method employed to prepare the initial states plays a primordial role. In this regard, the two primary approaches involve evolving a known quantum state via quantum techniques or directly loading a wave function obtained via classical methods. In the following, we will review some of the important methods to prepare initial states on a quantum computer. 

Adiabatic state preparation (ASP) methods have been widely studied to prepare ground-states~\cite{Albash_2018}. They encompass advancements such as shortcuts to adiabaticity \cite{STA}, counterdiabatic driving techniques \cite{CD_Berry,CD1,CD2}, or hybrid approaches like counterdiabatic optimal local driving \cite{COLD,barone2023counterdiabatic}. Additionally, VQE \cite{VQE_preuzzo}, \update{quantum imaginary time evolution \cite{motta2020qite}}, dissipative \cite{dissipation_cirac,dissipation_stefano, dissipation_ying, dissipation_zoller,motlagh2024ground} and \update{eigenstates filtering} \cite{Lin2020optimalpolynomial,Gluza2024doublebracket,robbiati2024} methods have some prospects. However, the scalability of these techniques to large system sizes remains to be determined. For instance, ASP demands an evolution time inversely proportional to the spectral gap, which can be exponentially small. Conversely, VQE is plagued by issues of trainability \cite{traps}, particularly concerning the barren plateau phenomenon \cite{cerezo2021cost,ragone2023unified} and high sample complexity \cite{PhysRevResearch.4.033154}. Finally, dissipative methods require coupling to a bath or transformations of the Hamiltonian using quantum signal processing, which can be expensive to implement.

The second category comprises techniques that load state vectors, given in a classical representation obtained by methods such as DMRG \cite{DMRG_White, DMRG_Wilson} and coupled cluster (CC) \cite{CC_review}, directly on the quantum computer. These techniques offer a crucial advantage by introducing cutoff parameters, such as the bond dimension $\chi$ for DMRG or the number of interacting orbitals for CC. Despite their steep scaling with system size, it is always feasible to identify an affordable cutoff that makes these methods implementable in practice. Generally, this yields a state far superior to a random guess or a mean-field state like Hartree-Fock, enabling the extraction of low\hyp lying eigenstate components using approaches like LT or QPE. An important step when using classical techniques is the loading of the obtained state onto the quantum computer. In fact, if the state vector is arbitrary, the cost is typically exponential in the number of qubits, e.g., using the scheme by \citet{mottononen}. Better scaling can be achieved if the state exhibits some special properties. For example, Refs~ \cite{tubman2018postponing, fomichev2024initial} developed a method to load multi\hyp Slater determinants and CC states, respectively. On the other hand, states obtained via DMRG can be loaded with overhead at most polynomial in the bond dimension \cite{loading_mps,loading_mps_cirac,Melnikov_MPS_QSP,smith2024constant,berry2024rapid}. As an extra step, we note that techniques have been introduced to further enhance initial state overlap, for example by optimizing the set of orbitals \cite{ollitrault2024enhancing}, or filtering out undesirable contributions \cite{Wang2022statepreparation}, while heuristics are provided in Ref.~\cite{gratsea2024comparing}.


We instead choose to address this challenge by sparsifying (or truncating) the classical wavefunction and using the sparse quantum state preparation proposed by \textcite{sparse_QSP}. The sparsification procedure consists of retaining only the $S$ largest components, setting all others to zero, and renormalizing. This $S$-sparsified state, denoted as $|\Psi_S\rangle$, can be efficiently loaded on the quantum computer, using only $\mathcal{O}(SN)$ two\hyp qubit gates, $\mathcal{O}(S\log{S}+N)$ single\hyp qubit gates \update{and no auxiliary qubit}, avoiding an exponential scaling. Even if DMRG states can be loaded with only a polynomial overhead in the bond dimension \cite{loading_mps,loading_mps_cirac,Melnikov_MPS_QSP,smith2024constant}, the sparsification procedure might be useful in the case that the prefactor is still significantly high. Moreover, sparsifying the state enables us to probe the regime where DMRG does not converge and serves as a proxy to study the LT algorithms would perform in this scenario. 

\subsection{Hamiltonian simulation}
\label{sec:ham-sim}
It's customary to quantify resource requirements in terms of calls to the time evolution oracle $\mathcal{U}(n\tau)$. However, a practical implementation necessitates an understanding of how to drive the dynamics efficiently. In order to keep the Heisenberg scaling, it is necessary to use asymptotically optimal methods based on linear combinations of unitaries (LCU) \cite{low_hamiltonian_2019, child_lcu}, generally scaling as $\mathcal{O}(\norm{\mathcal{H}}_1 \tau D + \log{\epsilon^{-1}})$ queries to the LCU decomposition. However, they come with significant drawbacks, such as requiring multiple ancillary qubits, high connectivity and a usual high pre\hyp factor. Therefore, we opt to utilize product formulas (PF) based on $p$\hyp order Trotter-Suzuki decomposition~\cite{SUZUKI1990319}, with an error scaling as $\epsilon \leq C (\tau D)^{p+1}/r^{p}$, and $r$ the number of Trotter steps and $C$ a prefactor depending on the commutators \cite{childs_theory_2021}, which can be determined specifically for each system, e.g. spins  \cite{PNAS_spins}, the Hubbard model \cite{campbell2021early} or neutrinos \cite{amitrano2023trapped}. This choice is motivated by two key factors: firstly, PFs do not entail ancilla overhead or costly control operations, and they often outperform what is theoretically guaranteed, even more when limited to the low-energy regime~\cite{hejazi2024better}. To contain the error below $\epsilon$, we can choose 
\begin{equation}
\label{eq:trotter_steps}
\begin{split}
r &= \left\lceil C^{1/p} (\tau D)^{(1+1/p)}\epsilon^{-1/p}\right\rceil.
\end{split}
\end{equation}
By doing so, we lose the Heisenberg scaling but have quantum circuits that are resilient to the restrictions of early FTQC devices. In fact, for $d$-local Hamiltonians, PFs can leverage vanishing commutators and achieve better scaling \update{in the system size} than using qubitization \cite{amitrano2023trapped}. Moreover, since the Fourier coefficients decay as $1/x$, we are more likely to run short-time simulations, which is the strong suit of PFs. This intuition is supported by our numerical experiments in Sec.~\ref{sec:res}, where good energy estimation is obtained even for circuits constrained to relatively low depths.

Although we do not explicitly delve into these techniques, various enhancements for product formulas exist, including but not limited to randomization \cite{Childs2019fasterquantum,doubling_random_berry, PhysRevA_random_knee}, multi-product formulas \cite{Faehrmann2022randomizingmulti, CarreraVazquez2023wellconditioned}, qDRIFT compilation \cite{QDrift,  concentration,qDRIFT_Kiss, nakaji_qswift_2023}, and composite formulas \cite{Rajput2022hybridizedmethods,Hagan2023compositequantum}. Since those techniques trade the circuit's depth with additional measurements, they are well suited for early FTQC, warranting further investigation in more targeted studies.

\subsection{Detection of discontinuities}
\label{sec:rpt}
The second main challenge we face is the detection of the discontinuities in the ACDF, which is directly related to the difficulty of computing a tight lower bound $\eta$ on the overlap with the true ground state. The problem is twofold: (i) having access to an upper bound $\eta$ that describes the size of the jump and is required to invert the CDF as described in \cite{Lin2022} is difficult, and (ii) since the number of jumps grows with the size of the support, the CDF becomes quasi-continuous, making any jump detection difficult. This can already be seen from Fig.~\ref{fig:ex}, where the CDF with eight qubits looks much smoother than with four. This is tightly related to the orthogonality catastrophe \cite{orthogonal_catastrophy, lee2023evaluating, louvet2023go}, stating that the overlap between a random initial state and any eigenstate vanishes exponentially fast for larger systems. We note that the continuous nature of the CDF is driven by the size of the support of the initial state in the eigenbasis. Hence, bound states, or more generally states which are sparse in this basis, will lead to clear step functions. However, since preparing initial states of this quality remains an open problem, it is important to consider states with exponential support. 


We adopt a different perspective, seeking to understand the practical implications of employing LT with an initial state of unknown quality.
Rather than striving for the exact determination of the ground state energy, we focus on enhancing the energy estimate of the initial state.
To this end, we aim to locate the inflection point of the CDF, \update{which we define as the first energy value which has a non-zero density with high probability. We argue that, in general, the inflection point improves over the energy expectation value of the initial state \cite{fomichev2024initial}. In fact, the expectation value corresponds to the mean energy of the state, while the inflection point corresponds to the low-energy part in the density of states. Moreover, since we make sure that the density is non-vanishing using statistical tests, we ensure that the inflection upper bounds the true ground-state energy with high enough probability.} We address this new task by identifying a statistically significant increase in the ACDF comprised of small contributions from neighbouring eigenstates. \reupdate{We note that, in essence, our proposal is an extension of the approach of \cite{Lin2022} when knowledge of the overlap is unavailable and aims to obtain the best estimate given a limited shot budget or runtime.}
\update{Intuitively, we are turning the question around and asking ourselves what the upper bound $E_{c(\eta)}$ is on our prediction within the available resources.  Given a maximal depth $D$, the minimal increase that can be detected is directly lower bounded by the approximation error of the Heaviside function $\epsilon_{D}$. We can then define the projector on the low\hyp energy regime 
\begin{equation}
\label{eq:accumulation}
\begin{split}
\mathcal{P}(\eta) &=\sum_{i=0}^{c(\eta)} |E_i\rangle \langle E_i|\quad \text{with} \\
c(\eta) &= \min_{n\in \mathbb{N}}{\sum_{i=0}^{n} |\langle \Psi|E_i\rangle \langle E_i| \Psi \rangle| \geq \eta>2\cdot\epsilon_{D}},
\end{split}
\end{equation}
where $c(\eta)$ is a truncation such that the overlap of the initial states in the low-energy manifold is higher than $\eta$. By turning the problem around, we are asking what is the best we can do given limited resources.  We emphasize that the accumulation is technically not required in the algorithm but instead serves as a pedagogical tool.}

We propose a procedure based on a kernel change-point detection method \cite{Celisse2018KCPD,Arlot2019KCPD}; this is a signal processing technique to detect changes in the mean of a given signal, implemented with the software package \textit{ruptures} \cite{rupture}.
Despite its simplicity, the technique can be used to perform complex time series analysis, yielding great results across multiple scientific domains \cite{Ambroise2019KCPDGenomics,Wang2021CPDCovid,Bringmann2022KCPDPsycho,Requena2023STEP}.
\reupdate{In this scenario, we execute an iterative procedure (Algorithm~\ref{alg:break}) that comprises three primary steps:
\begin{enumerate}[Step 1., font=\bfseries, leftmargin=1.5cm]
    \item  Dividing the signal into two segments, which entails identifying the breakpoint that optimally separates the time series.
    \item Evaluating the statistical significance of the split using the ANOVA-based scheme, with failure probability $\alpha_1$.
    \item Testing, with failure probability $\alpha_2$, that the jump is significant compared to the empirical noise.
\end{enumerate}
If these steps pass for a breakpoint, we repeat them on the lower-energy (left) part of the signal cut at the validated split. We thus find new breakpoints closer to the initial point with the lowest energy iteratively until one is rejected.}

\normalem 
\begin{algorithm}[!t]
\setstretch{1.2}

\caption{\textsc{Find smallest breakpoint}}
\label{alg:break}
\KwIn{signal $\{y_i\}_{i=1}^n$,
failure probabilities $\alpha_1, \alpha_2 \in[0,1]$}
\SetKwFunction{FindBP}{FindBP}
\SetKwFunction{ANOVA}{ANOVA}
\SetKwFunction{JUMP}{JUMP}
\nlnonumber \FindBP{y} \Comment*[r]{Finds breakpoint (Eq.~\ref{eq:rupture_min})}
\nlnonumber \JUMP{y, b, $\alpha_2$} \Comment*[r]{Validates jump (Thm.~\ref{thm:jump})}

$b \gets n$ \;
significant $\gets$ True \;
\While{\emph{significant}}{
$\tilde{b}$ = \FindBP{$y_{1:b}$} \Comment*[r]{Optimization}
significant = \ANOVA{$y$, $\tilde{b}$, $ \alpha_1$}  \rereupdate{\textbf{and} \JUMP{$y$, $\tilde{b}$, $\alpha_2$} \Comment*[r]{Validation using Thms.~\ref{thm:anova}-\ref{thm:jump}}}
\If{\emph{significant}}{
$b \gets \tilde{b}$ \Comment*[r]{Upon successful validation}}
}
\KwOut{$b$}
\end{algorithm}

To find the breakpoint, the signal \update{$y \in \mathbb{R}^n$} is first mapped onto a reproducing Hilbert space with $\phi: \mathbb{R} \rightarrow \mathbb{H}$, implicitly defined as $\phi(y) = K(y,\cdot)$. The kernel function $K$, typically chosen to be Gaussian, induces the metric on the Hilbert space. We then \update{embed the signal in the Hilbert space and} solve the following minimization problem to find the smallest breakpoint
\begin{equation}
    \label{eq:rupture_min}
    \min_{b \in \{1, \ldots, n\}}\ \sum_{t=1}^{b}\norm{\phi(y_t)-\bar{y}_{1:b}}+\sum_{t=b+1}^{n}\norm{\phi(y_t)-\bar{y}_{b+1:n}},
\end{equation}
where $\bar{y}_{a:b} = \sum_{x=a}^b y_x / (b-a) $ is the mean
of the signal between $a$ and $b$. \update{In the context of this paper, the signal is $y_t = C(t) = p(t)*\Theta(t)$, where $-\pi/2 < t<\pi/2$ belongs to the energy grid used to compute the initial estimate.} We guide the reader to Ref.~\cite{Celisse2018KCPD} for an in\hyp 
depth description of this technique.

\begin{algorithm}[t]
\SetAlgoLined
\setstretch{1.25}
\caption{\textsc{ANOVA}}
\label{alg:anova}
\KwIn{signal $\{y_i\}_{i=1}^n$,
breaking point $b\in[1,n]$,
failure probability $\alpha_1 \in[0,1]$} 
$\bar{y} = \frac{1}{n}\sum_{i=1}^n y_i$ \Comment*[r]{Mean of signal components}
$\bar{y}_0 = \frac{1}{b}\sum_{i=1}^b y_i$ \Comment*[r]{Mean of first segment}
$\bar{y}_1 = \frac{1}{n-b}\sum_{i\geq b} y_i$ \Comment*[r]{Mean of second segment}
$SS_{w} = \sum_{i=1}^{n} (y_i - \bar{y})^2$ \Comment*[r]{Sum of squares within segments}
$SS_b = \sum_{i<b} (y_i - \bar{y}_0)^2 + \sum_{i\geq b} (y_i - \bar{y}_1)^2$ \Comment*[r]{Sum of squares between segments}
$f = (n-2)SS_b / SS_w$ \Comment*[r]{F-ratio}
significant $\gets$ False \;
\If{$F(f,1,n-2) >(1-\alpha_1)$ \label{alg:f-test}}
{
\If{$\bar{y}_1>\bar{y}_0$}{
significant $\gets$ True \Comment*[r]{Upon successful F-test and check for monotonicity}
}}
\KwOut{significant $\in$ \{True, False\}}
\end{algorithm}
\ULforem

The ANOVA algorithm (Algorithm~\ref{alg:anova}) decides if the breakpoint is statically significant using a two-sample F-test and can be formally stated as:

\begin{theorem}[]
\label{thm:anova}
Let $X_i$ and $Y_i$ be the observations from two groups with sample sizes, \rereupdate{$n_X$ and $n_Y$, where} each group is drawn from a normal distribution with common variance $\sigma^2$ but possibly different means \rereupdate{$\mu_X \neq \mu_Y$}. \rereupdate{If the null-hypothesis, $H_0: \mu_X = \mu_Y$, is assumed to be true, then F-test (Line \ref{alg:f-test}, ANOVA) will reject it at a significance level $\alpha_1$, which represents the probability of incorrectly accepting a breakpoint.}

\end{theorem}
\begin{proof}
The $F$-statistic for them is given by
\rereupdate{
\begin{equation}
\begin{split}
    F &= (n-2) \frac{SS_b}{SS_w}, \quad \text{where}\ n = n_X + n_Y,\\
    SS_b  &=  \frac{n_Xn_Y}{n_X + n_Y} \left|\mu_X- \mu_Y\right|^2, \quad \text{and}\\
    SS_w  &= \sum_{i=1}^{n_X} (X_i - \mu_X)^2 + \sum_{j=1}^{n_Y} (Y_j - \mu_Y)^2.\\
\end{split}
\end{equation}
}
Under $H_0$, both groups have the same mean value, meaning the variation between groups is caused by pure randomness; we have that the $F$-statistic follows an $F$-distribution with degrees of freedom $(1, n- 2)$:
\begin{equation}
    F \sim F_{1, n - 2}.
\end{equation}
Since we assumed they follow a normal distribution, we have the following with $\chi^2_k$ as an eponymous distribution \cite{Kenney1951_chi2}
\begin{equation}
    \frac{SS_b}{\sigma^2} \sim \chi^2_{1}, \text{\quad and \quad}
    \frac{SS_w}{\sigma^2} \sim \chi^2_{n - 2},
\end{equation}
and their ratio follows an $F$-distribution, $F \sim F_{1, n - 2}$. By the definition, the probability of $F$ exceeding the critical value when $H_0$ is true is 
\begin{equation}
    P(F \geq F_{1,\,\rereupdate{n} - 2} \mid H_0\ \rereupdate{=\text{True}}) = \alpha_1.
\end{equation}
Thus, \rereupdate{assuming that $H_0$ is true, the test rejects it with probability $\alpha_1$}, completing the proof.
\end{proof}
One important assumption of the theorem \ref{thm:anova} is that the means are normally distributed. While this is likely the case when $M$ is large enough, this assumption can be formally tested using normality tests \cite{normality}.

\reupdate{Furthermore, we introduce a safeguard to avoid overshooting by stopping the procedure if the jump, defined as the difference in the mean of both segments, is smaller than $k \tilde{\sigma}$. Here $k \in \mathbb{N}$ defines the confidence level and $\tilde{\sigma}$ is the empirical standard deviation of the signal computed on the energy region \reupdate{$-\pi/2 \leq x<\pi/2$}, where no eigenvalues are present.}
\reupdate{
More rigorously, we have:
\begin{theorem}
\label{thm:jump}
Let \( \mu_X,\, \mu_Y \) be the means of two segments, and let \(\sigma\) be their common standard deviation. Suppose that the sample means \( \bar{X} \sim \mathcal{N}\left( \mu_X, \frac{\sigma^2}{n_X} \right) \) and \( \bar{Y} \sim \mathcal{N}\left( \mu_Y, \frac{\sigma^2}{n_Y} \right) \) are normally distributed. \rereupdate{If the null hypothesis, $H_0: \mu_X = \mu_Y$, is assumed to be true, then the} probability that a jump \( J \equiv \bar{X} - \bar{Y} \) is greater than \( k\sigma \), is $P(J > k\sigma \mid H_0) = 1 - \Phi(k)$,
where $\Phi(\cdot)$ is the CDF of the \rereupdate{standard} normal distribution \rereupdate{and $k \in \mathbb{N}$ is related to the probability ($\alpha_2$) of incorrectly accepting a jump as $k = \Phi^{-1}(\alpha_2)$}.
\end{theorem}

\begin{proof}
Under the null hypothesis, the random variables \( \bar{X} \) and \( \bar{Y} \) are independent, and their difference follows $J \sim \mathcal{N}\left( 0,\,\tilde{\sigma}^2 \right)$,
with 
\begin{equation}
\tilde{\sigma}^2 = \sigma^2 \left( \frac{1}{n_1} + \frac{1}{n_2} \right).
\end{equation}
Since we are interested in the probability \( P(J > k\tilde{\sigma} \mid H_0 = \text{True}) \), by standardizing \( J \) we obtain
\begin{equation}
P\left( \frac{J}{\tilde{\sigma}} > k \right)= P(Z > k),\quad k \in \mathbb{N}
\end{equation}
where \( Z \) follows a standard normal distribution that gives
$P(Z > k) = 1 - \Phi(k)$. This yields
\begin{equation}
P(J > k\sigma \mid H_0 = \text{True}) = 1 - \Phi(k) = 1 - \alpha_2,
\end{equation}
\rereupdate{with the user-defined probability $\alpha_2 = \Phi(k)$ being the probability of not rejecting $H_0$ when it is true.}
\end{proof}}
The total failure probability can be taken as the product of the ones of the two individual tests $\alpha=\alpha_1 \alpha_2$. To make the scheme more robust, we repeat the above procedure multiple times to compute the median of means using different data for the ACDF. Once the inflection point $\tilde{b}$ is known, the energy guess is chosen as the maxima of the gradient of the ACDF around a small window $\delta$ around the inflection point, $[\tilde{b}-\delta/2,\ \tilde{b}+\delta/2]$, as depicted in Fig.~\ref{fig:algo}. \reupdate{ While this does not give any theoretical guarantee on the precision, which would require the knowledge of the overlap $\eta$, this does ensure with a significant probability that the estimated energy from our approach is upper bounded by the one given by the vanilla LT approach.}


In summary, the signal is processed using ruptures to find a breakpoint, which is then validated according to a statistical test and comparing the jump to the statistical noise. If accepted, we perform the same analysis on the early part of the signal up to the breakpoint and discard the rest. We repeat these steps until a breakpoint is discarded and take the last valid breakpoint as a guess for the inflection point of the CDF. Since the eigenvalue is situated in the middle of the jump, it is important to look at the gradient of the CDF or use a Gaussian kernel, to improve the precision. The first significant maximum of the gradient exactly corresponds to an eigenvalue, while the secondary peaks are due to the approximation with the Fourier series. The whole procedure described in this section can be summed up as following steps - 
    \begin{enumerate}[Step 1., font=\bfseries, leftmargin=1.5cm]
        \item Prepare the best possible ground state using the available methods in your toolbox, e.g., DMRG ~\cite{DMRG_White}, coupled cluster (CC)~\cite{CC_review}, etc.
        \item Load the state on the quantum computer, e.g., by sparsification~\cite{sparse_QSP}, or by encoding them as sums of Slater determinants or matrix-product states (MPS)~\cite{fomichev2024initial}.
        \item Sample $M$ time indices from the coefficients of the Fourier series [Eq. \eqref{eq:fourier_coeffs1}].
        \item Use a product formula to compute one sample of the corresponding Fourier moment with a suitable number of steps [Eq. \eqref{eq:trotter_steps}].
        \item Build the approximate CDF (ACDF) using $M$ samples [Eq. \eqref{eq:gofx}].
        \item Identify the inflection point $x$ of the ACDF using \textit{ruptures} [Eq. \eqref{eq:rupture_min}].
        \item The energy estimate is the maxima of the ACDF’s derivative in a $\delta$-window around $x$.
    \end{enumerate}

\section{Resources estimates}
\label{sec:resources}

In light of the challenges that stem from implementing the LT algorithm in practice, especially obtaining tight lower bounds $\eta$, and the potentially small overlap between the initial state and the target ground state, we strive to find what is the best that can be done using limited resources. More precisely, we tackle the following problem.
\begin{problem}
\label{prob:prob2}
    Given a maximal depth $D$ and a maximal shot budget $M$, what is the best upper bound on the ground-state energy
    that can be provided by the algorithm?
\end{problem}
\update{This question was already answered in the previous section while introducing the notion of accumulation in Eq.~\eqref{eq:accumulation}. However, in this section, we provide more details about the prefactor and compute quantitative resource estimates of the LT algorithm}, which falls into two categories ---
the highest Fourier moment $D$, which is related to the maximal evolution time of the Hamiltonian simulation via $T_{\text{max}}=\tau D$, responsible for the bias in approximating the ACDF, and the number of samples $M$, which is associated with the variance of the output energy. We note that $D$ is directly linked to the maximal depth of the quantum circuits, which is why we also relate to it as maximal runtime.
\begin{theorem}
\label{th1}
    For a given accuracy $\epsilon \in (0, 1)$, the maximal runtime $D$ required to approximate the error function $\erf{(\cdot)}$ with $\epsilon$-error is given by
\begin{equation}
D = 2\cdot \left \lceil \sqrt{f\left(\beta,\, \min{\left[1,\ 4e^{-w(\epsilon) / 2}\right]} \right) \cdot w(\epsilon)}\ \right  \rceil + 1,
\end{equation}
where $\beta$ is chosen according to Eq. \eqref{eq:beta} and
\begin{equation}
\begin{split}
    w(\epsilon)&=W_0\left(\frac{18}{\pi \epsilon^2}\right), \\ f(\beta,\epsilon) &= - \frac{\ln{\epsilon} + \beta}{W_0\left(-\left[1 + \beta^{-1}\ln{\epsilon} \right] e^{-1}\right)}\ , \\
\end{split}
\end{equation}

with $W_0(\cdot)$ being the principal branch of the Lambert-W function.
\end{theorem}

\begin{proof}
The proof of this theorem can be based on  Ref.~\cite[\textit{Theorem 3}, Appendix A]{wan_randomized_2022}, which describes a Fourier approximation to the Heaviside function as $\max |\Theta(x) - F(x;\beta)| \leq (\epsilon_1 + \epsilon_2 + \epsilon_3) / 2 = \epsilon$ ~\cite[Eq. (A12)]{wan_randomized_2022}, where $\epsilon_{\{1,2\}}$ are the errors due to finite truncation $d$ and scaling parameter $\beta$, respectively, in approximating the error function using Eqs. (\ref{eq:fourier-series}, \ref{eq:fourier_coeffs1}), and $\epsilon_{3}$ is the error in approximating the Heaviside function with an error function.
For the choice $\epsilon_1=\epsilon_2=\epsilon_3=2\epsilon/3$ in this approximation, and using the values of $\beta$ and $d$ such that of $-(\epsilon_1 + \epsilon_2)/2 \leq F_{d}(x,\beta) \leq 1 + (\epsilon_1 + \epsilon_2)/2$ ~\cite[Eq. (A13)]{wan_randomized_2022}, we get $d\geq\sqrt{t\cdot w(\epsilon)}$~\cite[Eq. (A6)]{wan_randomized_2022} for an integer $t$ where
\begin{equation}
\label{eq:t-int}
 t =
  \begin{cases}
  f(\beta,\, 4 e^{-w(\epsilon) / 2}) & \text{for}\ w(\epsilon) \geq 2\ln4 \\
  \beta & \text{for}\ w(\epsilon) < 2\ln4 \\
  \end{cases},
\end{equation}
with the function $f(\cdot, \cdot)$ defined above. Since $f(\beta, 1) = \beta$ and $\epsilon\in (0, 1)$, one can rewrite Eq. \eqref{eq:t-int} as $t = f(\beta,\, \min\left[1,\, 4 e^{-w(\epsilon) / 2}\right])$ and obtain the minimum maximal runtime $D = 2d + 1$ as $2\left\lceil \sqrt{t\cdot w(\epsilon)}\,\right\rceil+1$.
\end{proof}

One can then use this maximal runtime for the estimation of the number of samples that are required to resolve an accumulation of size $\eta$. It is important to note that $\eta$ is no longer a lower bound on the overlap but a parameter chosen by the user, quantifying the desired resolution.
\begin{theorem}
\label{th2}
    For a given maximal runtime $D$ and accuracy $\epsilon \in (0, 1)$, the number of samples $M$ required to guarantee a correct result with probability $1-\vartheta$ is
    
    \begin{equation}
    \begin{split}
        M =& \Bigg\lceil 2 \cdot \left[\frac{2.07\pi^{-1} (\log{4D} + \gamma) + 1}{\eta - 2\epsilon}\right]^2 \cdot \\    
          & \left[\log{\log{\left(\frac{1}{\tau\epsilon}\right)} + \log{(\vartheta^{-1})}}\right] \Bigg\rceil,
    \end{split}
    \end{equation}
    
    where $\eta > 2\epsilon$ is the accumulation on the low\hyp energy part of the spectrum and $\gamma$ is the Euler-Mascheroni constant. The last assumption is required since a jump smaller than the error threshold cannot be resolved. 
\end{theorem}

\begin{proof}
We make use of $\mathbb{E}[G] = \tilde{C}(x)$ from Eq.~$\eqref{eq:acdf_def}$ to decide the Problem 1 from Eq.~$\eqref{eq:prob_1}$ as:
\begin{equation}
\begin{split}
G < \zeta\ & \Rightarrow\ \mbox{Tr}[\rho \Pi_{\leq x - \delta}] < \eta \\
G \geq \zeta\ &\Rightarrow\ \mbox{Tr}[\rho \Pi_{\leq x + \delta}] > 0.
\end{split}
\end{equation}

We use $\zeta = \eta/2$ to attune for errors in estimating $G(x)$ from sampling (Eq. \eqref{eq:gofx}) and decide whether $\tilde{C}(x) = 0$ or $\tilde{C}(x) \geq \eta$. Consequently, we can define the above as the probability $\mathbb{P}[G < \eta/2] < \kappa$ conditioned on $\tilde{C}(x) < \eta - \epsilon$ which implies $C(x - \delta) < \eta$, and use 
the one-sided Chebyshev inequality to find
\begin{equation}
\begin{split}
\mathbb{P}[G < \eta/2]&<\mathbb{P}[G \leq \eta/2]\\
&=\mathbb{P}[G \leq \tilde{C}-\eta/2+\epsilon]\\
&\leq \frac{\mbox{Var}[G]}{\mbox{Var}[G]+\left(\eta/2-\epsilon\right)^2}\\
&<\frac{\mbox{Var}[G]}{\left(\eta/2-\epsilon\right)^2}\;.
\end{split}
\end{equation}

Using the upper bound on the variance of $G$ discussed after Eq.~\eqref{eq:gofx} above we finally obtain

\begin{equation}
\mathbb{P}[G < \eta/2] < \frac{8\mathcal{F}^2}{M(\eta-2\epsilon)^2}\;,
\end{equation}

and in order to guarantee that this probability will be bounded by $\kappa$ we can then take

\begin{equation}
M\geq 
\left(\frac{2\sqrt{2}\mathcal{F}}{\eta-2\epsilon}\right)^2 \left(\frac{1}{\kappa}\right)\;.
\end{equation}

We can improve the dependence on $\kappa$ by using the fact that the individual summands in the average that defines $G$ are bounded by $2\mathcal{F}/M$ which allows us to employ the Hoeffding inequality as follows

\begin{equation}
\begin{split}
\mathbb{P}[\bar{G} < \eta/2] <& \exp\left(-\frac{2 \cdot \left(\eta/2 - \epsilon\right)^2}{M \cdot (2\mathcal{F}/ M)^2}\right)\\
&\Rightarrow M \geq 
\left(\frac{2\sqrt{2}\mathcal{F}}{\eta-2\epsilon}\right)^2 \log\left(\frac{1}{\kappa}\right)\;.
\end{split}
\end{equation}

We can now use the upper bound for the norm of the Fourier coefficients $\mathcal{F}$ computed as \cite{wan_randomized_2022}:

\begin{equation}
    \label{eq:eq_fval2}
    \begin{split}
    \mathcal{F} = \sum_{|j|\leq d} |\hat{F}_j(\beta)| &\leq \frac{2.07}{2\pi}(H_{D}+2\ln{2}) + \frac{1}{2}  \\&\approx \frac{2.07}{2\pi}\left(\log(4D) + \gamma  \right) + \frac{1}{2},
    \end{split}
\end{equation}

\begin{figure}[!t]
    \centering
    \includegraphics[width=\linewidth]{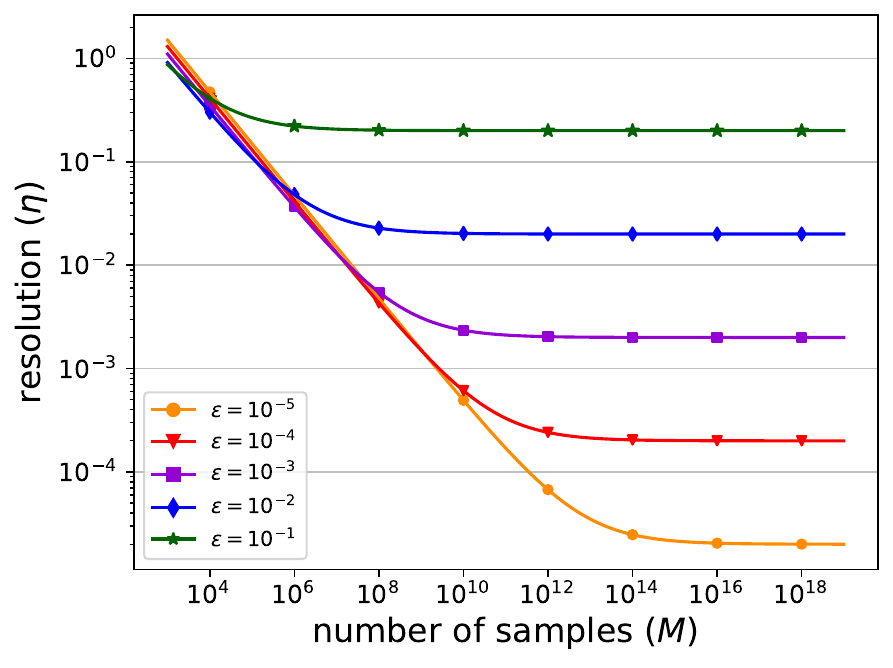}
    \caption{\justifying \textbf{Estimating the resolution.} Estimation of the resolution $\eta$ of a jump that can be resolved for the case of 26 spins fully connected Heisenberg model discussed in Sec. \ref{sec:large-sys}, using a given number of samples $M$ with $1-\vartheta = 95\%$ confidence for different precision parameters $\epsilon$.}
    \label{fig:samples}
\end{figure} 
where $H_k$ denotes the k$^{\text{th}}$ Harmonic number, which can be estimated as done above using the Euler-Mascheroni constant $\gamma = 0.57721567$ \cite{Graham1994-op}.
\begin{figure}[!t]
    \centering
    \includegraphics[width=\linewidth]{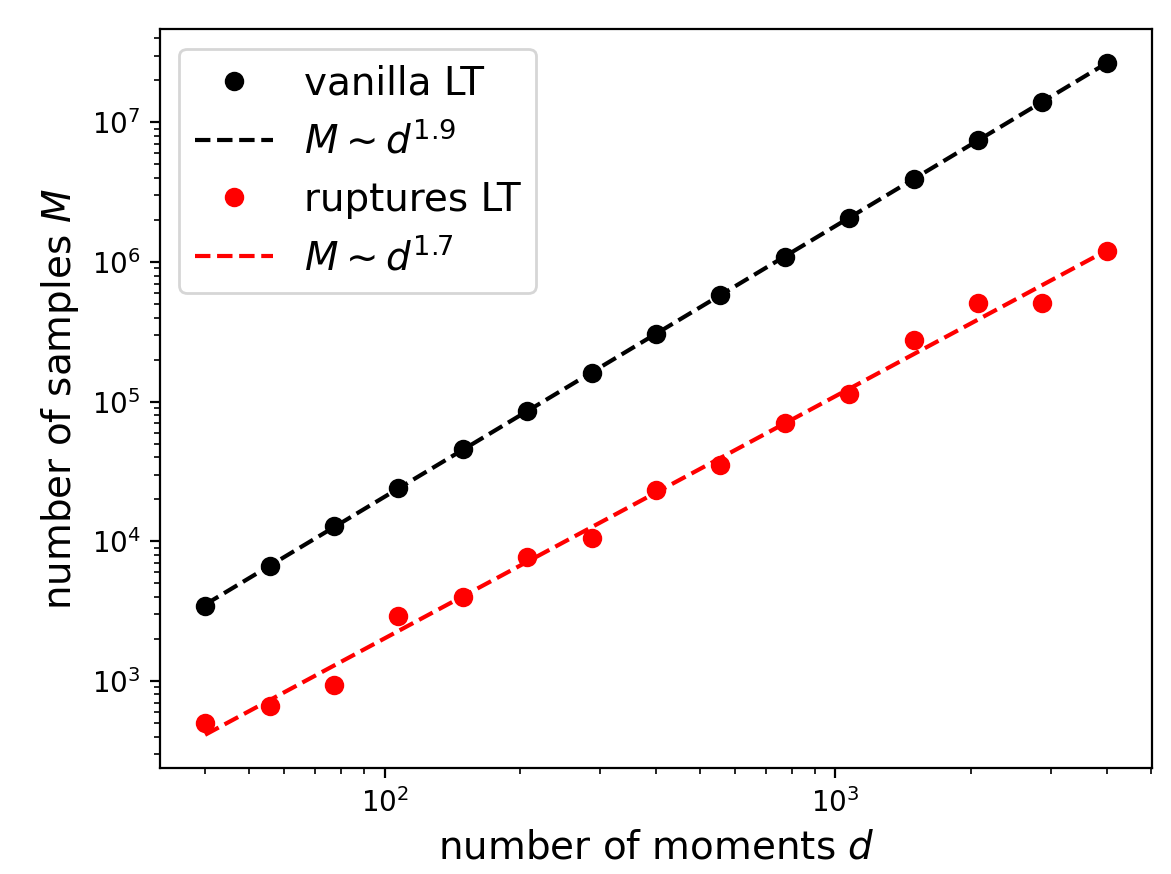}
    \caption{\textbf{Resources comparison.} \reupdate{Maximum} number of samples required to detect a step of size $\eta=2\epsilon_d$ for an artificial signal. The black dots show an \reupdate{upper bound} for the vanilla LT procedure using the resources estimates of Theorem \ref{th2}. The red dots are the empirical results using ruptures and ANOVA. \reupdate{The failure probabilities $\vartheta$ and $\alpha$ are $0.05$.}}
    \label{fig:comparison}
\end{figure}

Finally, we can get the provided result by noting that Algorithm~\ref{alg:break} needs to be run $\mathcal{O}(\log\delta^{-1})$ times to solve Problem 1 \cite{Lin2022}, and if it fails with probability at most $\kappa$ then to successfully estimate the ground state of the system with probability $1 - \vartheta$ we would have $\kappa^{-1} \leq \vartheta^{-1}\log\delta^{-1}$,
with $\delta=\tau\epsilon$ as before. 

\end{proof}
\begin{figure*}[!t]
    \centering
        \includegraphics[width=0.825\linewidth]{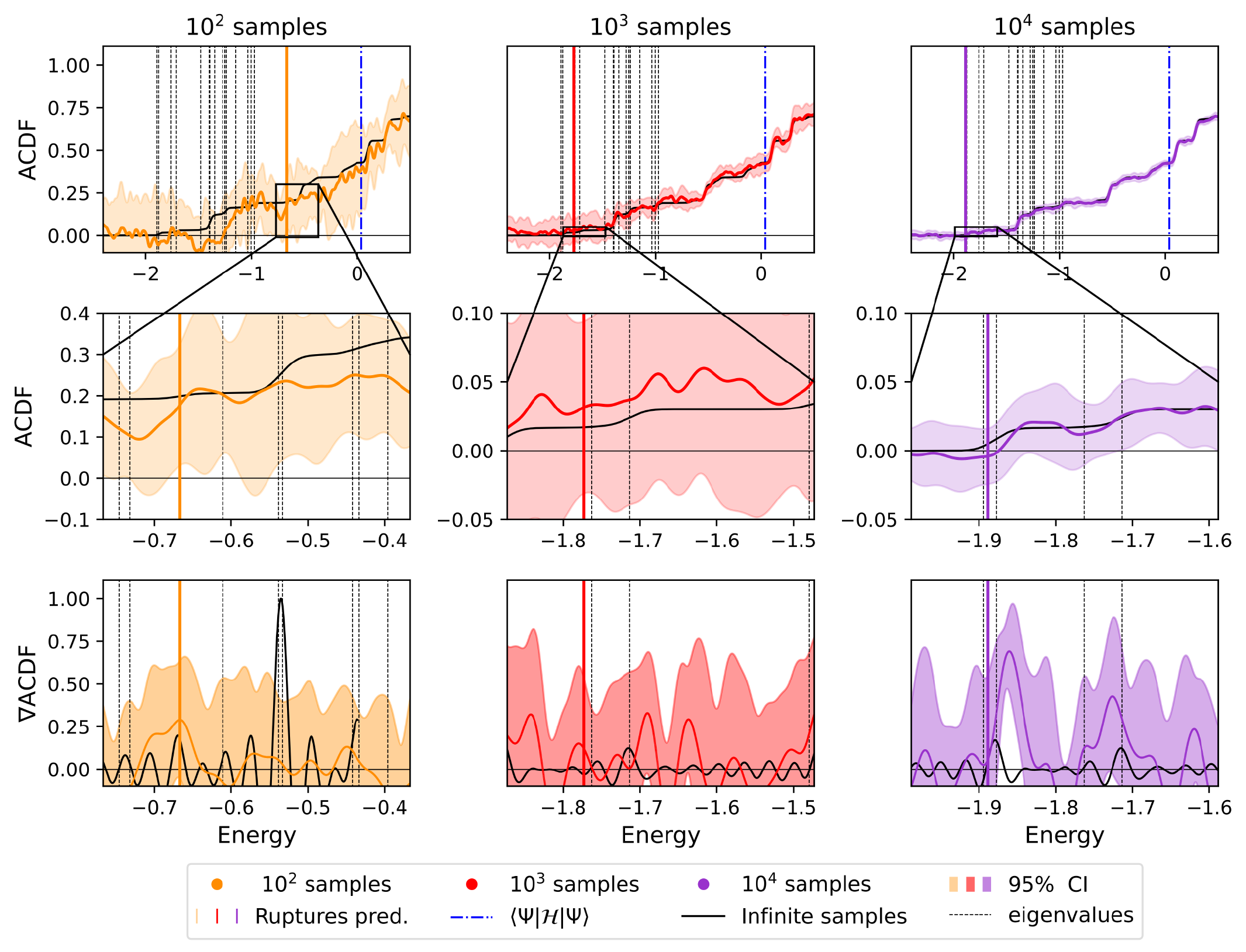}
    \caption{\justifying \textbf{Six spins, random initial state.}  ACDF (top row), zoom ACDF (middle row) and its gradient ($\nabla$ACDF, bottom row) computed with different number of samples (each column). The first 20 eigenvalues obtained from exact diagonalization are reported with dashed vertical lines, while the thick blue vertical line shows the energy of the initial state. The coloured line represents the energy found with the \textit{ruptures} procedure. The black line shows the ACDF (and its gradient) computed with infinite statistics, while the shaded area is a 95\% confidence interval computed with ten repetitions.}
    \label{fig:6spins}
\end{figure*}

\begin{figure*}[!t]
    \centering
    \includegraphics[width=0.825\linewidth]{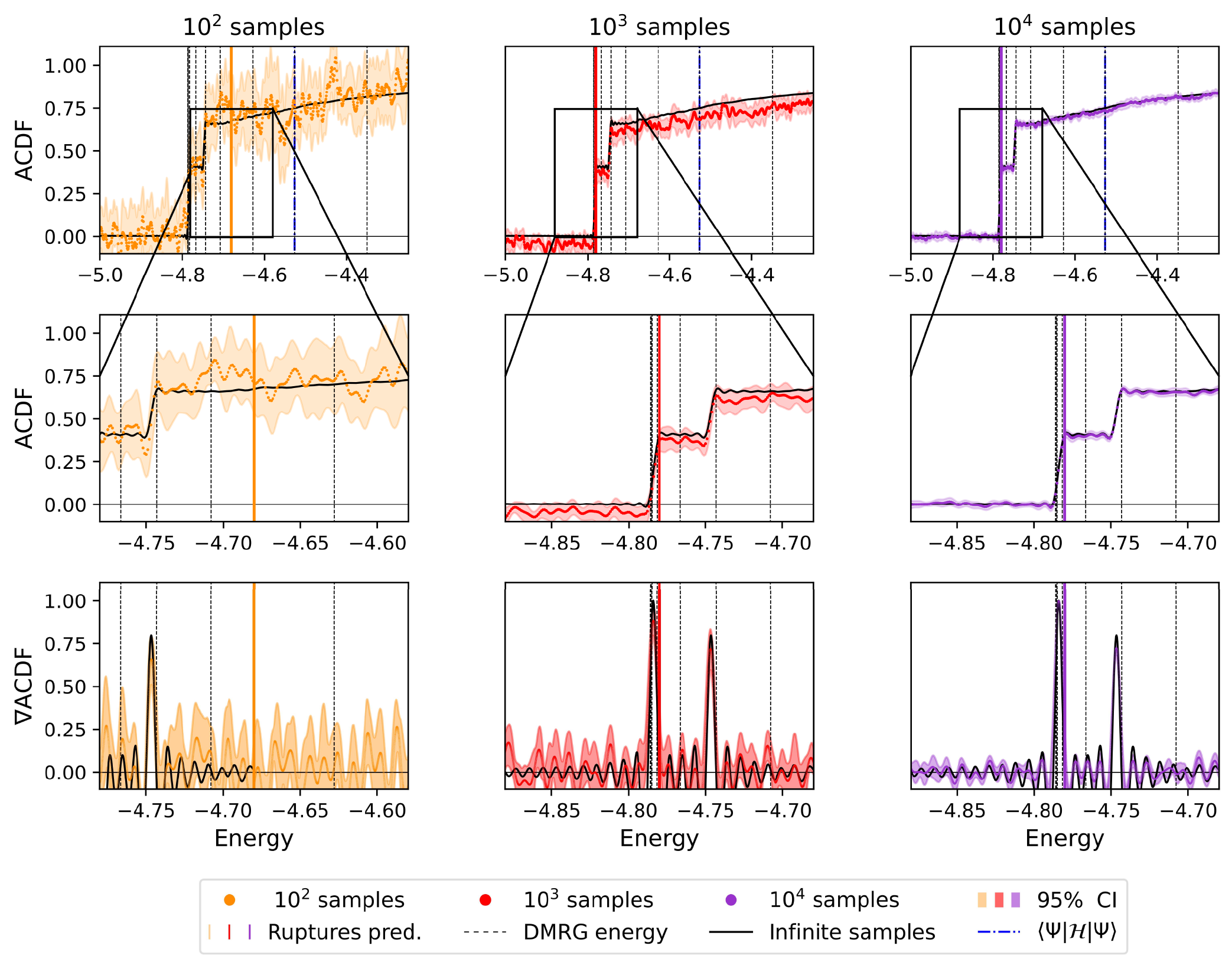}
    \caption{\justifying \textbf{26 spins, DMRG \update{initial state} ($\mathbf{\chi = 10}$).}  ACDF (top row), zoomed ACDF (middle row) and its gradient ($\nabla$ACDF, bottom row) computed with the different numbers of samples (each column). The energies obtained with untruncated DMRG states with different bond dimensions $5\leq \chi\leq 2000$ are reported with dashed vertical lines, while the thick blue vertical line shows the energy of the initial state. The coloured line represents the energy found with the \textit{ruptures} procedure. The black line shows the ACDF (and its gradient) computed with infinite statistics, while the shaded region is a 95\% confidence interval computed with ten repetitions. }
    \label{fig:26spins-normal}
\end{figure*}

It is important to note that these estimates are meant to be used in estimating the error tolerance and resolution which can be achieved using limited resources $D$ and $M$. For instance, we use the Theorem \ref{th2} in Fig.~\ref{fig:samples} to show the inflections of size $\eta$ that can be resolved for different error tolerances $\epsilon$ with a given sample budget $M$. We observe that the resolution which can be achieved plateaus after a certain sample budget, meaning that for further refined resolution, we need to increase the simulation time (i.e., decrease $\epsilon$). In particular, this means that we cannot trade depth with samples indefinitely and that to reach a higher resolution, it is required to increase the depth. As stated in Theorem~\ref{th1}, the maximum depth $D$ is primarily determined by the error $\epsilon$ when approximating the step function. \update{We leverage these resource estimates to determine the maximum number of samples required for step detection. Given a maximal number of moments \(d\), which can be determined from the maximum depth supported by the device, the minimal detectable step size is lower bounded by \(\eta = 2\epsilon_d\), accounting for the approximation error. \reupdate{An upper bound on} the number of samples required to achieve this level of precision is determined by Theorem~\ref{th2}. However, using the proposed signal processing technique, \textit{ruptures}, we demonstrate that significantly fewer samples are required. 

To validate this, we construct an artificial signal with two eigenvalues and a ground state overlap \(\eta = 2\epsilon_d\), and calculate the minimal number of samples required to find the first eigenvalue with precision \(\delta = 0.01\). To enhance the robustness, we employ a median-of-means estimator, incorporating its additional overhead into the total sample count. The results, presented in Fig.~\ref{fig:comparison}, illustrate that our automatic step detection procedure achieves \rereupdate{a ten fold samples reduction} over the vanilla approach. \reupdate{Moreover, our approach gives energy estimates that are upper bounded by the ones computed by the vanilla approach, which come with guarantees. Therefore, the key feature of our approach is to offer an agnostic way of estimating the first jump which is at least as good as the original's, while being significantly more efficient with the number of samples.}
} We will now delve deeper into this observation by conducting numerical simulations on a more challenging system.

\begin{figure*}[!t]
    \centering
        \includegraphics[width=0.82\linewidth]{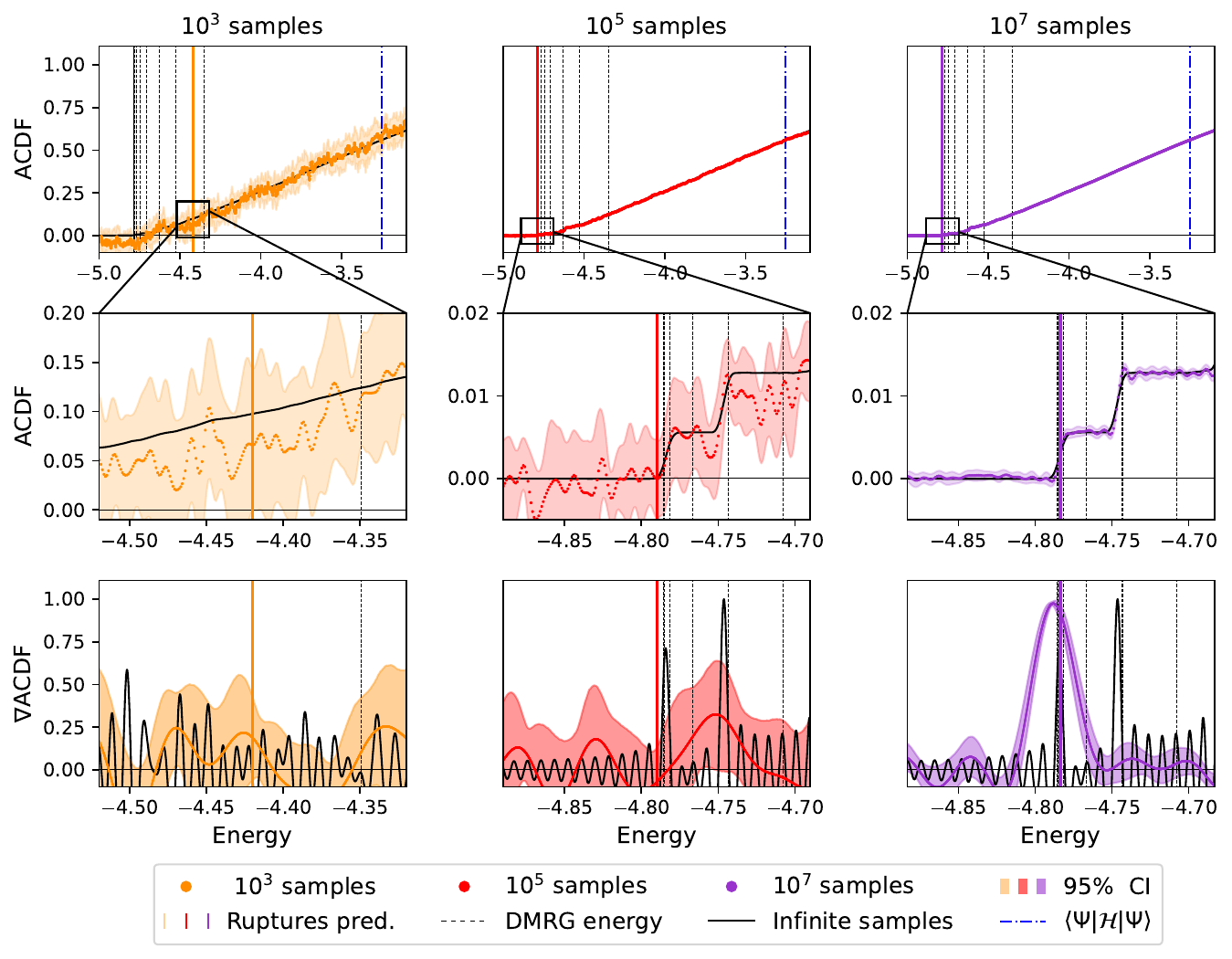}
    \caption{\justifying \textbf{26 spins, sparsified DMRG \update{initial state} ($\mathbf{\chi = 10}$).}  ACDF (top row), zoomed ACDF (middle row) and its gradient ($\nabla$ACDF, bottom row) computed with the different numbers of samples (each column). The energies obtained with untruncated DMRG states with different bond dimensions $5\leq \chi\leq 2000$ are reported with dashed vertical lines, while the thick blue vertical line shows the energy of the sparsified initial state with $S=13$. The coloured line represents the energy found with the \textit{ruptures} procedure. The black line shows the ACDF (and its gradient) computed with infinite statistics, while the shaded region is a 95\% confidence interval computed with ten repetitions.}
    \label{fig:26spins-sparse}
\end{figure*}

\section{Numerical simulations}
\label{sec:res}
We consider a fully connected Heisenberg model with random couplings over $N$ spins
\begin{equation}
\label{ham}
\begin{split}
\mathcal{H} &=\frac{1}{N} \sum_{i<j}  \sum_{a \in \{x,y,z\}}  J_a^{ij} \cdot \sigma_a^{i} \sigma_a^{j}, \text{ where } J_a^{ij} \sim \mathcal{N}(0,1).
\end{split}
\end{equation}
Here,  $\sigma_a^i$ is the corresponding Pauli matrix applied to the $i$-th qubit.
This model is gapless in general and universal, in the sense that it can approximate any two\hyp local Hamiltonian \cite{cubitt2018universal}. Moreover, the tensor network techniques are not expected to perform well due to the high number of connections, making it a good testbed for quantum algorithms. For this system, we simulate the dynamics using a second-order Trotter-Suzuki formula \cite{SUZUKI1990319} with a time step $\Delta t = \tau/8$, and a SWAP networks-based circuit construction \update{such that the circuit can be run with linear qubit connectivity \cite{swap_wiebe, swap_roggero}. The time step size is chosen such that the Trotter error is negligible compared to the error in the approximation of the Heaviside function. The state vector simulations are performed on GPUs hosted on a high-performance computing cluster.}

Using a product formula for approximating the time evolution has the advantage of keeping the circuit shallow, being, therefore, more suited for early FTQC devices. For instance, the depth of the quantum circuits used in the following reads $2NrD$. We note that we did not use any of the potential improvements to Hamiltonian simulation with product formulas discussed in Sec.~\ref{sec:ham-sim}, which are likely to decrease the maximal depth at the expense of additional samples. 


\subsection{Small system sizes}
We start with a small system with six spins ($N=6$). We choose a random initial state with $p_0=0.0014$ and $p_1=0.015$. Note that this is a state of poor quality, compared to DMRG. We also set $\epsilon = 0.055$ and therefore require $D=350$ Fourier moments for constructing the ACDF. 

Both the ACDF and its gradient are shown in Fig.~\ref{fig:6spins} for different numbers of samples. The continuous black line, referred to as the infinite statistics limit, corresponds to the infinite samples regime, where the moments are exact, and all of them, up to $D$, are included in the approximation. The ground-state energy is estimated with the procedure introduced in Sec. \ref{sec:rpt}, in which we first find an approximate guess for the inflection point and then refine it by taking the maximum of the gradient around it. We observe that we are not able to find the ground state, which is due to the approximation error of the step function $\epsilon$ being larger than the overlap. However, we are able to find the first excited-state energy using only $10^4$ shots and a random initial state. We note that we are able to find the energy of the true ground state using a better initial state, e.g. a DMRG state of low bond dimension.

\subsection{Large system sizes\label{sec:large-sys}}

We then move to a larger model composed of $N=26$ spins. Since exact diagonalization is too expensive, in this regime, we refer to DMRG with bond dimension $\chi=2000$ to obtain the target energy. We choose $\epsilon=0.019$, translating into $D=6600$. We perform two experiments: the first starting from $|\Psi\rangle = \text{DMRG}(\chi=10)$, and the second from its sparsification with $S=13$. The motivation to use DMRG with low bond dimension is to have a good initial state, which can be computed even for large system sizes.

The results with the DMRG state are shown in Fig.~\ref{fig:26spins-normal}, which displays the ACDF computed with $10^2$, $10^3$ and $10^4$ samples. The data displays the two main contributions from low-level energy eigenstates and a continuous contribution from higher states. With our procedure, we can recover the ground-state energy equivalent to DMRG with $\chi=2000$ from the gradient of the ACDF. An important point to make at this stage is that the CDF clearly exhibits two jumps on the low\hyp lying part of the spectrum. This is due to the high quality of the initial state, which consists of two bound states of low energy and a tail of high\hyp energy residuum. Therefore, the left part of the CDF exhibits a step-like shape, while the right part is continuous. However, as seen in Fig.~\ref{fig:26spins-sparse}, the CDF obtained from the sparsified DMRG state looks continuous, and we only unravel discontinuities when magnifying by a factor of one hundred. The important point is that in both cases, we improve on the DMRG estimate (blue dotted vertical line), by $5\%$ and $32\%$ respectively.

\begin{figure}[!t]
    \centering
    \includegraphics[width=\linewidth]{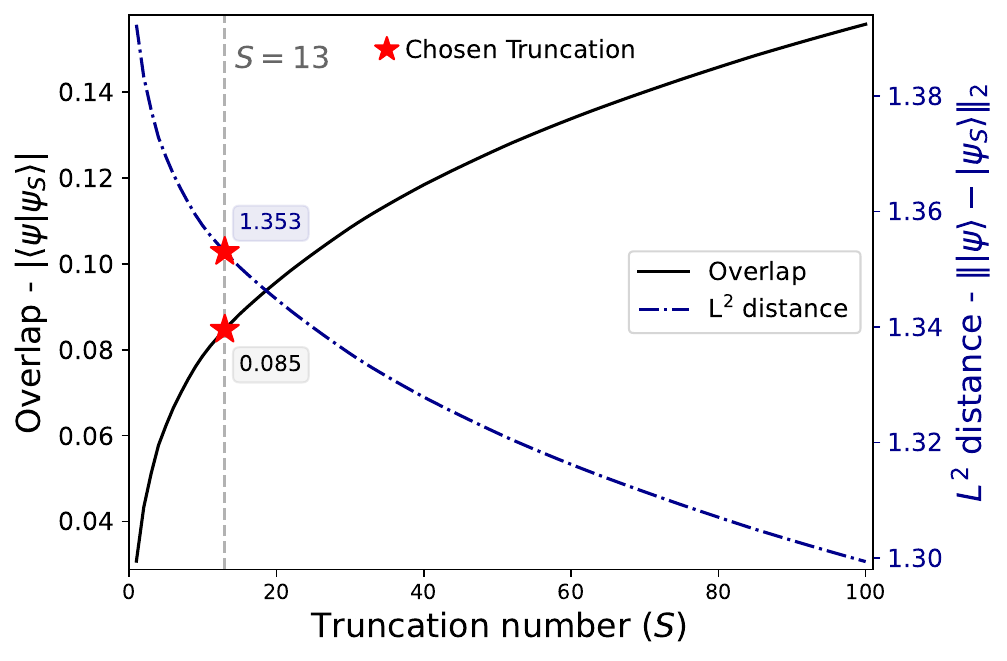}
    \caption{\justifying \textbf{Quality of the sparsification} The left axis (black) shows the overlap of $|\Psi_S\rangle$ with the full DMRG initial state, while the right axis (blue) shows the corresponding L$^2$ distance. The red stars correspond to the truncation number chosen for the experiments.}
    \label{fig:sparsity}
\end{figure}

The key difference with respect to the above case is that the ACDF does not display any clear steps and, as such, we have to rely on finding the inflection point. This results in the need of more samples to find the ground state energy, requiring $10^5$ and $10^7$ samples to recover the energy of DMRG with $\chi=1000$ and $\chi=2000$, respectively. To study the quality of the sparsified initial state $|\Psi_S\rangle$, we show the $L^2$ distance and the overlap with the true DMRG initial state as a function of the truncation parameter $k$ in Fig.~\ref{fig:sparsity}. We observe that the quality increase slows down after $S=13$, which is the reason this truncation was chosen.
Furthermore, we note that the overlap squared $p_0$ between the (sparsified) initial state and the converged DMRG state reads ($2\times 10^{-5}$) $7\times 10^{-6}$, requiring $M=10^{13}$ (using Theorem~\ref{th2}) samples to distinguish the exact ground state with $95\%$ certainty. This is why we should instead search for inflection points since it requires several orders of magnitude less resources than what is expected theoretically from detecting discontinuities while achieving similar performance. The important point here is the change of paradigm, going from aiming at the true ground\hyp state energy, which is challenging and costly, to detecting a relevant accumulation \update{that can be computed with much fewer resources while being useful for most applications.}

\begin{figure}

\centering
\subfloat[26 qubits, DMRG ($\chi=10$)]{%
  \includegraphics[clip, width=0.97\columnwidth]{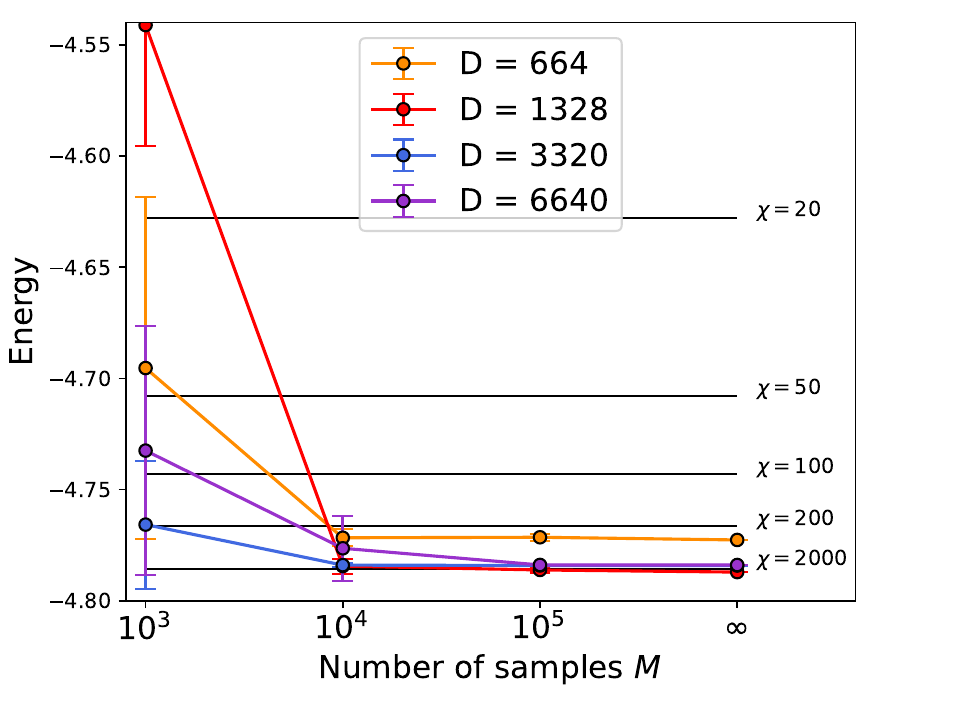}%
}\\
\subfloat[26 qubits, sparsified DMRG ($\chi=10$)]{%
  \includegraphics[clip, width=0.97\columnwidth]{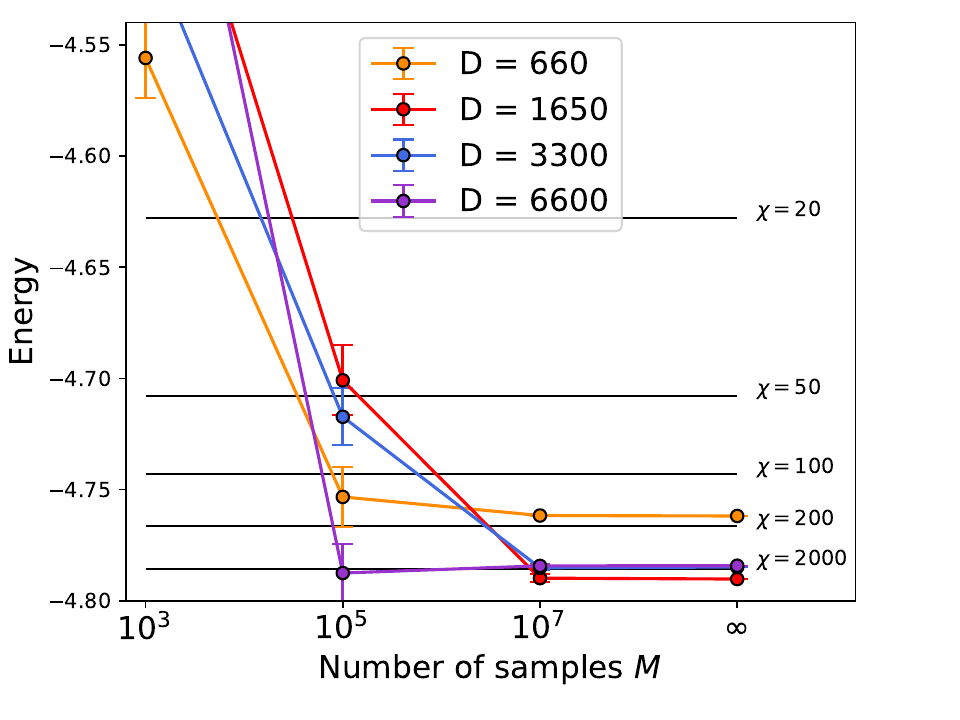}%
}

\caption{\justifying Energy predictions using the method introduced in Section \ref{sec:rpt} taking the maxima of the gradient of the ACDF in a $\delta=0.01$ window. The horizontal lines depict the values predicted by DMRG with different bond dimensions $\chi$. The different colours refer to the maximum time evolution used to compute the ACDF, while the x\hyp axis depicts the number of samples. The error bars represent one standard deviation computed over ten experiments. The infinity symbol refers to the regime of infinite samples.}
\label{energy_prediction}
\end{figure}

Finally, we show the energy predictions as a function of the number of samples and maximal depth in Fig.~\ref{energy_prediction}. We observe that with only a fifth of the depth considered in the previous experiments, we can already obtain the same energy estimates using just $10^3$ and $10^5$ samples, respectively, for the full and sparse initial states. This hints that finding the inflection point requires less precision than a jump, thus that fewer moments are required, in practice.

\section{Conclusion\label{sec:conc}}

In this paper, we examine the practical performance of early fault-tolerant algorithms for ground-state energy estimation. Specifically, our focus here has been on the Lin and Tong (LT) algorithm \cite{Lin2022}, which approximates the cumulative distribution function (CDF) of the spectral measure and associates its discontinuities with the eigenvalues of the corresponding Hamiltonian. Notably, this algorithm requires only one ancillary qubit and the ability to perform real-time evolution, making it a promising candidate for the intermediate-scale quantum hardware. We have summarized the current state\hyp of\hyp the\hyp art setup of this algorithm and make three distinct contributions there to enhance its practicality: the identification of the relevant challenges appearing in practice and how to address them; quantitative resource estimation in terms of maximal shot budget and evolution time for Hamiltonian simulation; and numerical simulation on a non\hyp trivial system. We advocate using product formulas for time evolution, even if we lose the Heisenberg scaling by doing so. Hence, they enjoy low implementation overhead and are likely better for the short-time evolution of local Hamiltonian, rather than the asymptotically optimal but more expensive techniques.

In light of the numerical simulation performed on the challenging fully\hyp connected, fully\hyp random Heisenberg model, LT\hyp type algorithms emerge as a viable option for early fault-tolerant quantum (FTQ) devices. The key point was to change our mindset from aiming at the actual ground\hyp state energy, which requires high depth and many samples, to instead concentrating on finding the spectral CDF's inflection point. \reupdate{Especially, when the initial state has a large accumulation but low overlap, \textit{ruptures} finds this point as the first statistically significant step, showing improvement over the original one. While in the case where the initial state is of high enough quality, as seems to be the case for a (sparsified) DMRG state of low bond dimension, the inflection point lies close enough to the true ground-state energy, our approach gives an estimate as good as the original approach if not better, which is usually precise for most applications.} Hence, this paradigm change has the advantage of relaxing the requirement in the approximation error, which is responsible for scaling the depth and number of samples. While quantum phase estimation is still likely outperforming the LT algorithm in the FTQ computing era, our findings suggest that LT algorithms can improve on classical solutions, using only limited quantum resources ($10^4-10^5$ samples), orders of magnitude less than what is predicted by the theory. In conclusion, the LT algorithm emerges as a robust and efficient quantum algorithm, bridging the gap between NISQ and FTQ computing eras.

\section*{Code} 
We use \texttt{Tenpy} \cite{tenpy} for the DMRG calculations, \texttt{PennyLane} \cite{2018arXiv181104968B, 2024arXiv240302512A} and \texttt{Qsim} \cite{quantum_ai_team_and_collaborators_2020_4023103} for the dynamics and \texttt{Ruptures} \cite{rupture} for the breakpoint detection. \update{The complete code is available on the following \href{https://github.com/XanaduAI/EFTQA-GSEE/tree/main}{GitHub repository}.} Numerical experiments were performed on the CERN Openlab GPU cluster.

\section*{Acknowledgments} 
The authors thank Stepan Fomichev and Francesco Di Marcantonio for useful discussions about DMRG and initial state preparation. This work was conducted while O.K. and B.R. were summer residents at Xanadu Inc. O.K. is additionally funded by the University of Geneva through a Doc. Mobility fellowship. A.R. is funded by the European Union under the Horizon Europe Program - Grant Agreement 101080086 — NeQST.
B. R. acknowledges support from Europea Research Council AdG NOQIA;
MCIN/AEI (PGC2018-0910.13039/501100011033, CEX2019-000910-S/10.13039/501100011033, Plan National FIDEUA PID2019-106901GB-I00, Plan National STAMEENA PID2022-139099NB, I00, project funded by MCIN/AEI/10.13039/501100011033 and by the “European Union NextGenerationEU/PRTR" (PRTR-C17.I1), FPI); QUANTERA MAQS PCI2019-111828-2); QUANTERA DYNAMITE PCI2022-132919, QuantERA II Programme co-funded by European Union’s Horizon 2020 program under Grant Agreement No 101017733);
Ministry for Digital Transformation and of Civil Service of the Spanish Government through the QUANTUM ENIA project call - Quantum Spain project, and by the European Union through the Recovery, Transformation and Resilience Plan - NextGenerationEU within the framework of the Digital Spain 2026 Agenda;
Fundaci\'o Cellex;
Fundaci\'o Mir-Puig;
Generalitat de Catalunya (European Social Fund FEDER and CERCA program, AGAUR Grant No. 2021 SGR 01452, QuantumCAT \ U16-011424, co-funded by ERDF Operational Program of Catalonia 2014-2020);
Barcelona Supercomputing Center MareNostrum (FI-2023-1-0013);
Funded by the European Union. Views and opinions expressed are however those of the author(s) only and do not necessarily reflect those of the European Union, European Commission, European Climate, Infrastructure and Environment Executive Agency (CINEA), or any other granting authority. Neither the European Union nor any granting authority can be held responsible for them (EU Quantum Flagship PASQuanS2.1, 101113690, EU Horizon 2020 FET-OPEN OPTOlogic, Grant No 899794), EU Horizon Europe Program (This project has received funding from the European Union’s Horizon Europe research and innovation program under grant agreement No 101080086 NeQSTGrant Agreement 101080086 — NeQST);
ICFO Internal “QuantumGaudi” project;
European Union’s Horizon 2020 program under the Marie Sklodowska-Curie grant agreement No 847648;
“La Caixa” Junior Leaders fellowships, La Caixa” Foundation (ID 100010434): CF/BQ/PR23/11980043.

\normalem

\bibliographystyle{unsrtnat}
\bibliography{Quantum/bibliography_quantum}

\end{document}